\documentclass[graybox, envcountchap]{svmult}

\usepackage[authoryear]{natbib}

\usepackage{comment}
\usepackage{siunitx}
\usepackage{newtxtext,newtxmath}

\usepackage{amsmath}
\usepackage{color}
\usepackage{helvet}          
\usepackage{courier}         
\usepackage{dirtree}
\usepackage{makeidx}        
\usepackage{graphicx}       
\usepackage{subfig}

\usepackage{multicol}        
\usepackage[bottom]{footmisc}

\usepackage{hyperref}        
\hypersetup{colorlinks=true,urlcolor=blue}
\hypersetup{colorlinks=true,urlcolor=blue} \makeatletter \providecommand*{\toclevel@title}{0} \providecommand*{\toclevel@titlech}{0} \providecommand*{\toclevel@author}{1} \providecommand*{\toclevel@authorch}{1} \providecommand*{\toclevel@mtchap}{1} \providecommand*{\toclevel@mtsec}{2} \makeatother \usepackage[misc]{ifsym} \makeindex

\usepackage[misc]{ifsym}

\makeindex             

\begin{document}


\title{Magnetohydrodynamics Simulations}
\author{Eliu Huerta}
\institute{Eliu Huerta (\Letter) \at Argonne National Laboratory, Lemont, IL 60439, USA \& The University of Chicago, Chicago, Illinois 60637, USA \& University of Illinois Urbana-Champaign, Urbana, IL 61801, USA \email{elihu@{anl.gov, uchicago.edu}}}
%
\maketitle

\abstract{Magnetohydrodynamics (MHD) couples the Navier--Stokes equations to Maxwell's equations, yielding a nonlinear system of partial differential equations (PDEs) that governs phenomena ranging from stellar interiors and astrophysical jets to fusion-reactor plasmas and space weather. Over the past half century, advances in numerical methods, from finite-volume Godunov schemes and constrained-transport algorithms to high-order spectral-element, discontinuous-Galerkin, and adaptive-mesh-refinement techniques, have transformed these equations from abstract theory into predictive tools, enabling high-fidelity simulations of solar eruptions, tokamak confinement, and magnetised turbulence. Yet a fundamental resolution barrier persists. In three-dimensional MHD turbulence the number of degrees of freedom required to resolve all dynamically active scales grows as $\mathcal{O}(\mathrm{Re}^{9/4})$ or steeper, where $\mathrm{Re}=LV/\nu$ is the fluid Reynolds number defined by a characteristic velocity~$V$, length scale~$L$, and kinematic viscosity~$\nu$. This scaling renders direct numerical simulation at astrophysically and fusion-relevant parameters intractable even on exascale platforms; in such regimes the Lundquist number $S=Lv_{A}/\eta$, the magnetic analogue of the Reynolds number formed from the Alfv\'{e}n speed~$v_{A}$ and the magnetic diffusivity~$\eta$, can exceed $10^{10}$, implying an even larger effective number of degrees of freedom when both viscous and resistive dissipation scales must be captured. Concurrently, the demand for multi-physics coupling, including kinetic closures, radiation transport, and uncertainty quantification, compounds the computational cost far beyond what traditional discretisations alone can sustain. This chapter reviews how artificial intelligence (AI) is opening a credible path to close this gap. We trace the progression from physics-informed neural networks (PINNs), which embed governing equations as soft constraints via automatic differentiation, through Fourier neural operators (FNOs) and their physics-informed extensions (PINOs), which learn solution operators across families of MHD problems, to hybrid operator-diffusion frameworks that combine deterministic operator surrogates with score-based generative models to recover the broadband spectral content of developed turbulence. Each methodological advance is driven by specific mathematical demands of the MHD equations, in particular the preservation of the solenoidal constraint $\nabla\cdot\mathbf{B}=0$ and the faithful representation of dual energy cascades across velocity and magnetic fields. We situate these AI methods within the broader computational landscape: exascale-ready high-order codes that exploit GPU acceleration and task-based parallelism to push resolution limits; data-driven sub-grid-scale closures that reduce grid requirements while preserving cascade dynamics; and emerging quantum algorithms for sparse linear-system solves that may, as fault-tolerant hardware matures, offer additional reductions in time-to-solution for the implicit systems central to resistive MHD. Although each component is at a different stage of maturity, their synergistic integration, physics-constrained network architectures embedded within conventional solvers, trained on leadership-scale computing facilities, and potentially accelerated by quantum kernels, charts a compelling trajectory for the field. The chapter aims to equip the reader with both the theoretical foundations of MHD turbulence and a critical assessment of the AI tools poised to extend the reach of MHD simulation into regimes that have remained, until now, beyond computational access.}


\section{Introduction and Historical Context}
\label{sec:1}

The governing equations of MHD have been shaped over more than eight decades through three overlapping epochs. In the foundational plasma‑physics era of the 1940s–1960s, Hannes Alfvén recognised that a perfectly conducting fluid supports transverse electromagnetic disturbances, now known as Alfvén waves \citep{Alfven1942}. Cowling subsequently proved the impossibility of maintaining an axisymmetric magnetic field by axisymmetric motions alone, thereby introducing magnetic diffusion as a central concept~\citep{Cowling1934,Cowling1957}. The single‑fluid description that couples the Navier–Stokes equations with Maxwell’s equations was consolidated in the classic monographs of Braginskii \citep{Braginskii1965} and Kulsrud \citep{Kulsrud2005}.

During the 1970s–1990s the focus shifted to stability and astrophysical synthesis. Chandrasekhar’s treatise presented the energy principle and systematic normal‑mode analyses of magnetised instabilities \citep{Chandrasekhar1961}. Parker's $\alpha$--$\omega$ dynamo theory demonstrated how the interplay of differential rotation ($\omega$-effect) and cyclonic turbulence ($\alpha$-effect) can systematically amplify seed magnetic fields to the strengths observed on stellar and galactic scales~\citep{Parker1979}. In the fusion community, ideal‑MHD equilibrium theory was codified by Freidberg \citep{Freidberg1987}, while Goedbloed and Poedts provided a unified treatment of ideal and resistive MHD \citep{GoedbloedPoedts2004}, cementing the equations as predictive tools for tokamaks and stellarators.

The advent of high‑performance computing in the 2000s inaugurated the computational and multi‑physics era. Godunov-type upwind schemes together with constrained-transport (CT) methods enabled the preservation of the divergence-free condition $\nabla\cdot\mathbf{B}=0$ in three-dimensional simulations. The CT algorithm was introduced by Evans \& Hawley~\cite{EH_CTA} and subsequently implemented within the ZEUS code by Stone \& Norman~\cite{StoneNorman1992}. CT maintains $\nabla\cdot\mathbf{B}=0$ to machine precision by evolving the magnetic field components on a staggered mesh: the magnetic flux through each cell face is updated using the electromotive force computed at cell edges, ensuring that Stokes' theorem is satisfied discretely. This geometric property makes CT the method of choice for long-duration MHD simulations where even small monopole errors can accumulate and corrupt the solution.

High‑order reconstruction \citep{Balsara1998}, the HLLD Riemann solver \citep{Miyoshi2005}, and hyperbolic divergence cleaning \citep{Dedner2002} further improved robustness for supersonic, highly magnetised turbulence. Community codes such as ZEUS \citep{StoneNorman1992}, ATHENA \citep{Stone2008}, PLUTO \citep{Mignone2007}, and the relativistic extension GR-Athena++ \citep{Cook_2025} have democratized access to large‑scale MHD modelling. 

\section{MHD in Numerical Relativity and General‑Relativistic Astrophysics}
\label{sec:grmhd}

In the past decade general‑relativistic magnetohydrodynamics (GRMHD) \citep{Gammie2003} has become a core tool for modelling compact‑object mergers, accretion onto black holes, and jet formation. Giacomazzo and Rezzolla presented one of the first three‑dimensional GRMHD simulations of magnetised neutron stars, establishing a benchmark for subsequent codes \citep{GiacomazzoRezzolla2007}. The IllinoisGRMHD code, released within the Einstein Toolkit, combines high‑order shock capturing with constrained‑transport and has been applied to binary neutron‑star merger simulations \citep{Etienne2015,Armengol2022}. Porth and collaborators introduced the \texttt{BHAC} code, which implements a staggered‑mesh CT scheme on adaptive mesh-refinement grids and has been used for global simulations of black‑hole accretion flows \citep{Porth2017}. Radiative GRMHD has been incorporated in \texttt{KORAL}, which solves the coupled radiation‑MHD equations with M1 closure \citep{Sadowski:2013gua}. The extension of Athena++ to GRMHD provides a versatile platform for both Newtonian and relativistic problems, offering sub‑second runtimes on modern GPU‑accelerated nodes \citep{Cook_2025}. These advances have been essential for interpreting multimessenger observations such as GW170817 and for predicting electromagnetic counterparts of compact‑object mergers. The convergence of robust divergence‑free discretisations, adaptive mesh refinement, and efficient relativistic Riemann solvers now permits fully three‑dimensional, high‑resolution GRMHD simulations on emerging exascale systems.

\section{Engineering Applications: Fusion and Fission}
\label{sec:engineering}

Non‑linear MHD modelling is a cornerstone of ITER (International Thermonuclear Experimental Reactor) and next‑step tokamak design. The \texttt{JOREK} code, a reduced‑MHD/full‑MHD hybrid, has been used to predict edge‑localized mode dynamics and disruption‑mitigation strategies for ITER \citep{Hoelzl2021,Artola2022}. Parallel efforts with the \texttt{M3D‑C1} code have demonstrated three‑dimensional resistive‑wall‑mode evolution and resonant magnetic perturbation optimisation \citep{Krebs2017}. Recent investigations of liquid‑metal plasma‑facing components employ high‑fidelity MHD simulations to assess turbulent heat transport and magnetic drag \citep{Sun_2023,Mistrangelo2021}.

In liquid‑metal fast reactors the interaction between the conductive Na/K coolant and imposed magnetic fields governs flow stability and heat removal. High‑resolution simulations using spectral‑element and finite‑volume solvers have quantified the suppression of turbulence by strong transverse fields \citep{YU2023112175}. For molten‑salt reactors, magnetic‑field effects on salt convection have been examined in the context of in‑core electromagnetic pumps \citep{NOWAK2022109142}. These studies inform the design of safety‑critical components such as intermediate heat exchangers, where accurate prediction of magnetically‑induced pressure drops is essential.

\section{Why AI--Coupled HPC is the Next Paradigm}
\label{sec:ai_transition}

Traditional high performance computing (HPC) solvers rely on explicit discretisation of the governing equations and on the repeated solution of massive linear systems. As the ambition of MHD modelling expands to ever larger Reynolds numbers, tighter scale separation, and richer physics (kinetic closures, radiation transport, neutronics), the required degrees of freedom exceed the limits of even exascale machines. This ``resolution gap'' manifests as either prohibitive wall‑clock times or uncontrolled model reduction, both of which jeopardise predictive capability for the most demanding scientific and engineering problems. AI techniques, whether embedded within a conventional solver or deployed as stand alone surrogates, offer a systematic route to close this gap.

Physics‑informed neural networks (PINNs) have been shown to solve forward and inverse MHD problems while honouring the divergence‑free constraint, without the need for a mesh‑based discretisation \citep{RAISSI2019686,jaillon2026physics}. Because the governing equations appear as soft constraints in the loss function, PINNs can seamlessly incorporate experimental or simulation data, enabling data‑driven discovery of sub‑grid closure terms that are otherwise inaccessible to analytic modelling \citep{Karniadakis2021}.  

Operator‑learning frameworks such as the Fourier neural operator (FNO) learn a mapping from physical parameters (e.g., resistivity, magnetic Prandtl number) to the full spatio‑temporal solution operator, providing orders‑of‑magnitude speed‑ups compared with traditional time‑stepping while preserving spectral accuracy \citep{Li2020}. These methods 
have been demonstrated on a variety of PDEs, including the Burger's equation, 
Darcy Flow, and Navier-Stokes \citep{Li2020,Rosofsky2023MLST}. More challenging 
applications involving 2D incompressible MHD equations have been presented in 
\citep{Rosofsky2023PINO} for Reynolds numbers below 500, and in \citep{Kacmaz2025} for 
very turbulent magnetized flows with Reynolds numbers up to 10,000.

Data‑driven subgrid‑scale models trained on high‑resolution direct numerical simulation data can replace explicit eddy‑viscosity terms in large‑eddy simulations, reducing the required grid resolution while retaining the correct cascade dynamics \citep{2021PhFl...33c1702S}. Because the learned model is differentiable, it integrates naturally with adjoint based optimisation and gradient‑based design, opening the door to automated control of fusion plasma scenarios and to inverse design of magnetic‑confinement configurations.  

In summary, the convergence of mature AI methodologies with the unprecedented parallelism of exascale architectures provides a credible path to transcend the resolution barrier that limits current MHD practice. By coupling data‑driven inference, operator learning, and physics‑preserving neural architectures with traditional solvers, the community can achieve both the fidelity required for scientific discovery and the speed needed for engineering decision‑making. The remainder of this manuscript explores concrete algorithmic strategies for such AI‑coupled HPC workflows.

\subsection{Magnetohydrodynamics: From theoretical foundations to numerical breakthroughs}

MHD turbulence governs the dynamics of many astrophysical plasmas, including
the interstellar medium, the solar wind, and magnetized accretion flows. In these
systems the Reynolds and magnetic Reynolds numbers are extremely large, allowing nonlinear interactions to transfer energy across a wide range of spatial scales.
Understanding this cascade of energy is fundamental to explaining a variety of astrophysical processes such as cosmic-ray scattering, magnetic field amplification,
and plasma heating.

The modern theory of MHD turbulence emerged through a sequence of conceptual and computational developments. Early theoretical work extended the phenomenology of hydrodynamic turbulence to magnetized fluids by considering the interaction of Alfv\'{e}n waves. Subsequent studies demonstrated that the presence
of a large-scale magnetic field introduces strong anisotropy into the cascade and
modifies the nonlinear interaction between fluctuations. These ideas were eventually validated and refined through large-scale numerical simulations, which have
revealed complex phenomena including alignment effects, intermittency, and imbalanced cascades.

This section reviews the physical foundations of MHD turbulence and describes
how theoretical models, numerical simulations, and modern computational methods
have shaped our understanding of turbulent astrophysical plasmas.

\subsection{Magnetohydrodynamic Equations}

The macroscopic dynamics of a conducting fluid coupled to the magnetic field through the Lorentz force is described by the equations of magnetohydrodynamics. In the resistive regime the governing equations, written in Gaussian CGS units, consist of the conservation of mass, conservation of momentum, the induction equation for the magnetic field, the energy equation, and the solenoidal constraint on the magnetic field.

\paragraph{Mass conservation.} 
\begin{equation}
\frac{\partial \rho}{\partial t} + \nabla \cdot (\rho\,\mathbf{v}) = 0\,,
\label{eq:mass}
\end{equation}
where $\rho(\mathbf{x},t)$ is the mass density and $\mathbf{v}(\mathbf{x},t)$ is the plasma velocity field.

\paragraph{Momentum equation.}
\begin{equation}
\rho \left( \frac{\partial \mathbf{v}}{\partial t} + \mathbf{v} \cdot \nabla \mathbf{v} \right)
=
-\nabla P
+
\frac{1}{4\pi}
\left( \nabla \times \mathbf{B} \right) \times \mathbf{B}
+
\rho\,\nu\,\nabla^{2}\mathbf{v}\,,
\label{eq:momentum}
\end{equation}
where $P$ is the thermal pressure, $\mathbf{B}$ is the magnetic field, and $\nu$ is the kinematic viscosity. The Lorentz force per unit volume,
\[
\frac{1}{4\pi}(\nabla \times \mathbf{B}) \times \mathbf{B}\,,
\]
can be decomposed into a magnetic pressure gradient
\[
-\nabla\left(\frac{B^{2}}{8\pi}\right)
\]
and a magnetic tension
\[
\frac{1}{4\pi}(\mathbf{B}\cdot\nabla)\mathbf{B}\,.
\]
In the ideal limit ($\nu \to 0$) the viscous term vanishes.

\paragraph{Induction equation.}
\begin{equation}
\frac{\partial \mathbf{B}}{\partial t}
=
\nabla \times (\mathbf{v} \times \mathbf{B})
+
\eta\,\nabla^{2}\mathbf{B}\,,
\label{eq:induction}
\end{equation}
where
\[
\eta = \frac{c^{2}}{4\pi\sigma}
\]
is the magnetic diffusivity, $\sigma$ is the electrical conductivity, and $c$ is the speed of light. In the ideal limit ($\eta \to 0$) the magnetic field is frozen into the plasma flow.

\paragraph{Energy equation.}
\begin{equation}
\frac{\partial e}{\partial t}
+
\nabla \cdot
\left[
\left(
e + P + \frac{B^{2}}{8\pi}
\right)\mathbf{v}
-
\frac{1}{4\pi}\,\mathbf{B}\,(\mathbf{v}\cdot\mathbf{B})
\right]
=
0\,,
\label{eq:energy}
\end{equation}
where the total energy density is
\begin{equation}
e
=
\frac{1}{2}\rho v^{2}
+
\frac{P}{\gamma - 1}
+
\frac{B^{2}}{8\pi}\,,
\label{eq:energy_density}
\end{equation}
and $\gamma$ is the adiabatic index. Dissipative source terms (ohmic heating, viscous dissipation) have been omitted for brevity but are straightforward to include. For isothermal or barotropic flows the system may alternatively be closed by specifying $P = P(\rho)$ directly.

\paragraph{Solenoidal constraint.}
\begin{equation}
\nabla \cdot \mathbf{B} = 0\,,
\label{eq:divB}
\end{equation}
which expresses the absence of magnetic monopoles. This constraint is not an evolution equation but a condition that must be satisfied at all times; its numerical preservation is a central challenge in computational MHD and a recurring theme of this chapter.

\paragraph{Alfv\'en velocity and field normalisation.} The magnetic field is often expressed in velocity units through

\begin{equation}
\mathbf{b}
=
\frac{\mathbf{B}}{\sqrt{4\pi\rho}}\,,
\label{eq:b_normalised}
\end{equation}
where the factor of $\sqrt{4\pi}$ arises from the Gaussian CGS unit system adopted throughout this chapter. A key parameter in magnetised plasmas is the Alfv\'en velocity
\begin{equation}
v_{A}
=
\frac{B_{0}}{\sqrt{4\pi\rho}}\,,
\label{eq:alfven_speed}
\end{equation}
which represents the propagation speed of Alfv\'en waves along the background magnetic field~$\mathbf{B}_{0}$.

\paragraph{Characteristic wave speeds.}
The ideal MHD system supports seven characteristic wave families. The fast and slow magnetosonic speeds are given by
\begin{equation}
c_{f,s}^{\,2}
=
\frac{1}{2}
\left[
\left(a^{2} + v_{A}^{2}\right)
\pm
\sqrt{
\left(a^{2} + v_{A}^{2}\right)^{2}
-
4\,a^{2}\,v_{A,n}^{2}
}
\right],
\label{eq:fast_slow}
\end{equation}
where
\[
a = \sqrt{\frac{\gamma P}{\rho}}
\]
is the adiabatic sound speed,
\[
v_{A} = \frac{B}{\sqrt{4\pi\rho}}
\]
is the total Alfv\'en speed, and
\[
v_{A,n} = \frac{B_{n}}{\sqrt{4\pi\rho}}
\]
is its component normal to the wave front. The remaining characteristic speeds are the shear Alfv\'en speed $c_{A} = v_{A,n}$ and the entropy wave speed $c_{e} = v_{n}$ (the normal component of the flow velocity). These seven wave families, namely, fast and slow magnetosonic modes, shear Alfv\'en waves, and the entropy wave, govern the rich phenomenology of shocks, rarefactions, rotational discontinuities, and turbulent cascades that MHD simulations must capture.

\medskip

These equations express the fundamental coupling between plasma motion and magnetic fields. The velocity field stretches and advects magnetic field lines, while magnetic tension and pressure feed back on the plasma motion through the Lorentz force. The dimensionless parameters that control the relative importance of advection, diffusion, and magnetic effects are the Reynolds number
\[
\mathrm{Re} = \frac{LV}{\nu}\,,
\]
the magnetic Reynolds number
\[
\mathrm{Rm} = \frac{LV}{\eta}\,,
\]
and the Lundquist number
\[
S = \frac{L v_{A}}{\eta}\,,
\]
where $L$ and $V$ are characteristic length and velocity scales. The magnetic Prandtl number
\[
\mathrm{Pr}_{m}
=
\frac{\nu}{\eta}
=
\frac{\mathrm{Rm}}{\mathrm{Re}}
\]
measures the ratio of viscous to resistive dissipation scales and varies by many orders of magnitude across physical systems: $\mathrm{Pr}_{m} \gg 1$ in the interstellar medium, $
\mathrm{Pr}_{m} \sim 1$ in parts of the solar convection zone, 
and $\mathrm{Pr}_{m} \ll 1$ in liquid-metal experiments and stellar interiors.

\subsection{Elsässer Representation}

We now specialize to the incompressible limit with uniform density $\rho = \rho_0 = \mathrm{const}$, which is appropriate when the flow speed is much smaller than both the sound speed and the Alfv\'en speed. 

In the incompressible limit the energy equation decouples from 
the system: the pressure is no longer a thermodynamic 
variable but a Lagrange multiplier determined instantaneously by the incompressibility constraint $\nabla \cdot \mathbf{v} = 0$ through a Poisson equation. The total pressure, $P$, appearing below $P = p/\rho_0 + b^2/2$, where $p/\rho_0$ is the thermodynamic pressure divided by the (constant) mass density, commonly termed the \emph{kinematic pressure}, and $b^2/2$ is the magnetic pressure expressed in Alfv\'{e}n-velocity units. Both contributions carry dimensions of velocity squared, consistent with the per-unit-mass form of the momentum equation in the incompressible limit. In this limit, the continuity equation reduces to the incompressibility constraint, and the momentum and induction equations simplify considerably. The dynamics of a perfectly conducting incompressible plasma 
are governed by the ideal magnetohydrodynamic equations

\begin{equation}
\nabla \cdot \mathbf{v} = 0 ,
\end{equation}

\begin{equation}
\nabla \cdot \mathbf{b} = 0 ,
\end{equation}

\begin{equation}
\frac{\partial \mathbf{v}}{\partial t}
+
(\mathbf{v}\cdot\nabla)\mathbf{v}
=
-\nabla P
+
(\mathbf{b}\cdot\nabla)\mathbf{b},
\end{equation}

\begin{equation}
\frac{\partial \mathbf{b}}{\partial t}
+
(\mathbf{v}\cdot\nabla)\mathbf{b}
=
(\mathbf{b}\cdot\nabla)\mathbf{v},
\end{equation}

where, as mentioned above, $\mathbf{v}$ is the velocity field and 
$\mathbf{b}=\mathbf{B}/\sqrt{4\pi\rho}$ is the magnetic field expressed in
Alfvén velocity units, the scalar $P$ denotes the total pressure including
both hydrodynamic and magnetic contributions. It is now convenient to introduce the Elsässer variables

\begin{equation}
\mathbf{z}^{\pm} = \mathbf{v} \pm \mathbf{b}.
\end{equation}

Using these variables the ideal MHD equations can be written in symmetric
form

\begin{equation}
\frac{\partial \mathbf{z}^{\pm}}{\partial t}
+
\left( \mathbf{z}^{\mp} \cdot \nabla \right)\mathbf{z}^{\pm}
=
-\nabla P
+
\nu_{+}\,\nabla^{2}\mathbf{z}^{\pm}
+
\nu_{-}\,\nabla^{2}\mathbf{z}^{\mp}\,,
\label{eq:elsasser_dissipative}
\end{equation}

where $\nu_{+} = (\nu + \eta)/2$ and $\nu_{-} = (\nu - \eta)/2$, with $\nu$ the kinematic viscosity and $\eta$ the magnetic diffusivity. When $\nu = \eta$ (i.e., $\mathrm{Pr}_{m} = 1$), the cross-coupling term vanishes and each Els\"asser field is damped independently. These equations reveal an important property 
of MHD dynamics:
nonlinear interactions occur only between oppositely propagating
Elsässer fields.

\subsection{Projection onto the Solenoidal Subspace}

The pressure term enforces incompressibility. It can be eliminated by
introducing the projection operator onto divergence–free vector fields,

\begin{equation}
\mathcal{S}_{ij} = \delta_{ij} - \frac{\partial_i\,\partial_j}{\nabla^{2}}\,.
\label{eq:leray}
\end{equation}

Here $\partial_{i}\partial_{j}/\nabla^{2}$ denotes the composition of the partial derivatives $\partial_{i}\partial_{j}$ with the inverse Laplacian $\nabla^{-2}$, defined such that $\nabla^{2}(\nabla^{-2} f) = f$ for any function $f$ with zero spatial mean. In Fourier space this operator acts as simple multiplication: 

\begin{equation}
\hat{\mathcal{S}}_{ij}(\mathbf{k}) = \delta_{ij}-
\frac{k_i k_j}{|\mathbf{k}|^{2}}\,.
\end{equation}

where $\Delta = \nabla^2$ denotes the Laplacian.
Applying this projection removes the pressure gradient, yielding

\begin{equation}
\partial_t \mathbf{z}^{\pm}
=
-
\mathcal{S}\left[(\mathbf{z}^{\mp}\cdot\nabla)\mathbf{z}^{\pm}\right].
\end{equation}

This form is commonly used in spectral simulations of incompressible
MHD turbulence because it preserves the divergence-free condition
without explicitly computing the pressure.

\subsection{Guide Field and Alfvén Propagation}

Astrophysical plasmas often contain a strong uniform magnetic field.
The magnetic field can therefore be decomposed as

\begin{equation}
\mathbf{b} = \mathbf{v}_A + \delta \mathbf{b},
\end{equation}

where

\begin{equation}
\mathbf{v}_A = \frac{\mathbf{B}_0}{\sqrt{4\pi\rho}}
\end{equation}

is the Alfvén velocity associated with the background field
$\mathbf{B}_0$. Substituting this decomposition into the Elsässer equations produces linear propagation terms,

\begin{equation}
\left(
\partial_t \mp \mathbf{v}_A \cdot \nabla
\right)
\mathbf{z}^{\pm}
=
-
S\left[(\mathbf{z}^{\mp}\cdot\nabla)\mathbf{z}^{\pm}\right].
\end{equation}

These terms represent Alfvén wave packets propagating along the
background magnetic field. The fields $\mathbf{z}^+$ and $\mathbf{z}^-$ represent Alfvénic perturbations propagating parallel and anti-parallel to the mean magnetic field. Nonlinear interactions occur only between oppositely directed wave packets, a property that fundamentally shapes the cascade dynamics. The energies associated with the Elsässer fields are

\begin{equation}
E^\pm = \frac{1}{2}\langle |\mathbf{z}^\pm|^2 \rangle ,
\end{equation}

and the difference between these energies corresponds to the cross-helicity of the system. In the absence of forcing and dissipation these quantities are conserved invariants of the ideal MHD equations.

\subsection{Reduced Magnetohydrodynamics}

In the presence of a strong guide field turbulent fluctuations become
highly anisotropic. Observations and simulations indicate the ordering

\begin{equation}
k_\perp \gg k_\parallel ,
\end{equation}

where $k_\perp$ and $k_\parallel$ denote wavenumbers perpendicular and
parallel to the mean magnetic field. Under this anisotropic ordering the MHD equations reduce to the
system of reduced magnetohydrodynamics (RMHD),

\begin{equation}
\left(
\partial_t \mp v_A \partial_\parallel
\right)
\mathbf{z}^{\pm}_\perp
+
(\mathbf{z}^{\mp}_\perp \cdot \nabla_\perp)
\mathbf{z}^{\pm}_\perp
=
-\nabla_\perp P .
\end{equation}

The RMHD equations retain the essential nonlinear dynamics of
Alfvénic turbulence while removing compressive fluctuations and
parallel components of the fields. They therefore form the
theoretical and computational foundation of modern studies of
magnetized turbulence.

\subsection{Energy Cascade and Critical Balance}

In turbulent plasmas energy injected at large scales cascades toward smaller scales through nonlinear interactions. In the presence of a strong mean magnetic field this cascade becomes highly anisotropic. The nonlinear interaction time at scale $\lambda_\perp$ is

\begin{equation}
\tau_{nl} \sim \frac{\lambda_\perp}{\delta v_\lambda},
\end{equation}

where $\delta v_\lambda$ is the characteristic velocity fluctuation at that scale. Alfvén waves propagate along magnetic field lines with a characteristic time

\begin{equation}
\tau_A \sim \frac{\lambda_\parallel}{v_A},
\end{equation}

where $\lambda_\parallel$ is the scale measured along the mean magnetic field. The critical balance hypothesis states that these two timescales become comparable in strong MHD turbulence,

\begin{equation}
\tau_{nl} \sim \tau_A.
\end{equation}

Combining the critical balance condition
\begin{equation}
\frac{\lambda_{\perp}}{\delta v_{\lambda_{\perp}}}
\sim
\frac{\lambda_{\parallel}}{v_{A}}
\end{equation}
with the Kolmogorov scaling
\begin{equation}
\delta v_{\lambda_{\perp}}
\sim
\varepsilon^{1/3}\,\lambda_{\perp}^{1/3}
\end{equation}
gives
\begin{equation}
\lambda_{\parallel}
\sim
\frac{v_{A}}{\varepsilon^{1/3}}\,\lambda_{\perp}^{2/3}\,.
\end{equation}

Using the estimate
\begin{equation}
\varepsilon \sim \frac{v_{A}^{3}}{L}
\end{equation}
for the energy cascade rate at the outer scale $L$, this simplifies to
\begin{equation}
\lambda_{\parallel}
\sim
L^{1/3}\,\lambda_{\perp}^{2/3}\,,
\label{eq:anisotropy}
\end{equation}
showing that fluctuations become increasingly elongated along the magnetic field at smaller perpendicular scales. This leads to a scale-dependent anisotropy between parallel and perpendicular directions,

\begin{equation}
\lambda_\parallel \propto \lambda_\perp^{2/3}.
\end{equation}

The critical balance hypothesis, introduced by Goldreich \& Sridhar states~\cite{1995ApJ...438..763G} that the energy spectrum of the turbulent cascade follows the Kolmogorov-type scaling

\begin{equation}
E(k_\perp) = C_K \varepsilon^{2/3} k_\perp^{-5/3},
\end{equation}

where $\varepsilon$ is the energy dissipation rate and $k_\perp$ is the perpendicular wavenumber. Numerical simulations confirm that the inertial-range spectrum and anisotropy of Alfvénic turbulence are controlled by the parameters $v_A$, $\varepsilon$, and the spatial scale $\lambda$ \citep{beresnyak2019}.

\subsection{Dynamic Alignment}

A key development in modern turbulence theory is the concept of dynamic alignment. In this picture the velocity and magnetic fluctuations become progressively aligned as the cascade proceeds to smaller scales. Boldyrev proposed that the alignment angle between velocity and magnetic fluctuations obeys~\cite{boldyrev2006}

\begin{equation}
\theta_\lambda \propto \lambda^{1/4}.
\end{equation}

This alignment weakens nonlinear interactions and leads to a modified energy spectrum

\begin{equation}
E(k_\perp) \propto k_\perp^{-3/2}.
\end{equation}

Alignment can be measured using the amplitude-weighted quantity

\begin{equation}
DA = 
\frac{\langle |\delta \mathbf{v}_\lambda \times \delta \mathbf{b}_\lambda| \rangle}
{\langle |\delta v_\lambda| |\delta b_\lambda| \rangle}\,,
\end{equation}

where $\delta \mathbf{v}_\lambda$ and $\delta \mathbf{b}_\lambda$ denote field increments across a spatial separation $\lambda$, and $\langle \cdot \rangle$ denotes an appropriate spatial or ensemble average. Numerical studies indicate that the scale dependence of this alignment is weaker than originally predicted, suggesting that alignment effects may be non-universal and possibly tied to outer-scale conditions \citep{mason2006}.

\subsection{Imbalanced Turbulence}

In many astrophysical plasmas the amplitudes of the two Elsässer fields are not equal. This regime is known as imbalanced turbulence. Imbalance occurs when

\begin{equation}
z^{+}_{rms} > z^{-}_{rms},
\end{equation}

meaning that fluctuations propagating in one direction dominate those propagating in the opposite direction. Such conditions are commonly observed in the solar wind, where outward-propagating Alfvén waves originate from the solar corona.

In imbalanced turbulence the cascade rates of the two Elsässer components differ, leading to asymmetric energy transfer and modified anisotropy relations. The dynamics of this regime remain an active area of research \citep{lithwick2007,beresnyak2008}.

\subsection{Intermittency and Structure Functions}

Turbulent energy dissipation is not uniformly distributed in space but instead occurs in localized structures such as current sheets. This phenomenon is known as intermittency. Intermittency is commonly quantified using structure functions

\begin{equation}
S_{p}(\lambda) = \left\langle \left| \delta w(\lambda) \right|^{p}
\right\rangle\,,
\end{equation}

where

\begin{equation}
\delta w(\lambda) = \left[
\mathbf{z}^{\pm}(\mathbf{x} + \boldsymbol{\lambda})
- \mathbf{z}^{\pm}(\mathbf{x}) \right]
\cdot
\hat{\boldsymbol{\lambda}}\,,
\end{equation}

is the longitudinal increment of an Els\"asser field, and $\langle \cdot \rangle$ denotes a spatial or ensemble average. Higher-order structure functions reveal deviations from simple power-law scaling, reflecting the presence of intermittent structures in the turbulent flow.

\subsection{Recent Developments}

Recent theoretical work suggests that aligned turbulent structures may become unstable to magnetic reconnection at sufficiently small scales. This leads to the concept of a tearing-mediated turbulent cascade, in which current sheets formed by the cascade are disrupted by tearing instabilities. Such processes may play a crucial role in determining the dissipation scale of magnetized turbulence.

Modern numerical simulations have reached extremely high resolutions, enabling detailed studies of anisotropy, intermittency, and alignment effects. These simulations produce enormous datasets describing the structure of turbulent plasmas across many scales. The growing availability of such datasets has opened new opportunities for data-driven modeling approaches, including machine learning methods designed to extract statistical relationships from turbulent flows, this is the theme of the 
following section.


\section{The Computational Landscape}  

In conservation form, the ideal MHD equations constitute a hyperbolic system supporting seven characteristic wave families,  fast and slow magnetosonic modes, shear Alfv\'{e}n waves, and an entropy wave, whose interactions produce a rich phenomenology of shocks, rarefactions, rotational discontinuities, and turbulent cascades~\citep{Brio1988,Toth2000}. 

Numerical solution of the MHD equations has a long and distinguished history. Finite-volume Godunov-type schemes, constrained transport algorithms, and high-order spectral methods have been implemented in community codes such as \textsc{Athena}~\citep{Stone2008}, \textsc{FLASH}~\citep{Fryxell2000}, and \textsc{Pluto}~\citep{Mignone2007}, enabling landmark simulations of MHD turbulence, magnetic reconnection, and plasma instabilities. Yet these methods face a fundamental computational barrier: the cost of direct numerical simulation (DNS) scales steeply with the Reynolds number $Re$ and the Lundquist number $S = Lv_A/\eta$, where $\eta$ is the magnetic diffusivity (equivalently, the resistivity divided by $\mu_0$ in SI units). The Lundquist number measures the ratio of the resistive diffusion time $L^2/\eta$ to the Alfv\'en crossing time $L/v_A$. 

In three-dimensional MHD turbulence, the number of degrees of freedom required to resolve all dynamically active scales grows as $\mathcal{O}(Re^{9/4})$ in the hydrodynamic Kolmogorov estimate; for MHD at magnetic Prandtl number $\mathrm{Pr}_m = \nu/\eta \neq 1$, the scaling is modified because the resistive and viscous dissipation scales differ, and the effective resolution requirement can be even more severe~\cite{beresnyak2019}, rendering fully resolved simulations at astrophysically or fusion-relevant parameters ($S \gtrsim 10^{10}$) intractable even on exascale platforms. This computational gap has motivated a sustained search for methods that can either accelerate existing solvers or learn approximate solution maps at dramatically reduced cost. Over the past five years, AI and machine learning (ML) have emerged as serious contenders for this role. The developments reviewed here trace a coherent intellectual arc: from physics-informed neural networks that solve individual PDE instances, through neural operators that learn solution maps across families of problems, to generative models that capture the stochastic structure of turbulence. Each stage addressed specific limitations of its predecessor, and each drew essential motivation from the particular mathematical structure of the MHD equations.   

\section{Physics-Informed Neural Networks} 

The foundational contribution that opened the modern era of AI-driven PDE solvers was the PINN framework of Raissi, Perdikaris, and Karniadakis~\citep{RAISSI2019686}. The idea is conceptually simple but powerful: a neural network $\mathcal{N}_{\boldsymbol{\omega}}(\mathbf{x},t)$, parameterized by trainable weights $\boldsymbol{\omega}$, approximates the solution $u(\mathbf{x},t)$ of a PDE. Rather than training the network solely on labeled data, one constructs a loss function that penalizes violations of the governing equations at a set of collocation points $\{(\mathbf{x}_i, t_i)\}_{i=1}^{N_r}$ sampled throughout the computational domain:

\begin{equation}
\begin{aligned}
\mathcal{L}(\boldsymbol{\omega})
&=
\frac{\lambda_r}{N_r}\sum_{i=1}^{N_r}
\left|
\mathcal{F}\!\left[\mathcal{N}_{\boldsymbol{\omega}}\right](\mathbf{x}_i,t_i)
\right|^2
+
\frac{\lambda_b}{N_b}\sum_{j=1}^{N_b}
\left|
\mathcal{B}\!\left[\mathcal{N}_{\boldsymbol{\omega}}\right](\mathbf{x}_j,t_j)
\right|^2
\\[0.5em]
&\quad+
\frac{\lambda_0}{N_0}\sum_{k=1}^{N_0}
\left|
\mathcal{N}_{\boldsymbol{\omega}}(\mathbf{x}_k,0)
-
u_0(\mathbf{x}_k)
\right|^2\,.
\end{aligned}
\label{eq:pinn_loss}
\end{equation}

The coefficients $\lambda_r$, $\lambda_b$, and $\lambda_0$ are hyperparameters that control the relative weighting of the PDE residual, boundary condition, and initial condition losses, respectively. Their selection significantly affects training dynamics and is itself an active area of research~\cite{Wang2021}. 
$\mathcal{F}[u]=0$ is the PDE, $\mathcal{B}[u]=0$ encodes boundary conditions, $u_0$ specifies initial data. 

All derivatives appearing in $\mathcal{F}$ are computed exactly via automatic differentiation through the network's computational graph, eliminating the need for finite-difference stencils or mesh generation. The appeal of PINNs for MHD is immediate. The solenoidal constraint $\nabla\cdot\mathbf{B}=0$, a persistent source of numerical difficulty in grid-based MHD codes, where violations can generate spurious parallel forces~\citep{Toth2000}, 
can be enforced architecturally by representing the magnetic field as the curl of a learned vector potential, $\mathbf{B} = \nabla\times\mathbf{A}$, which satisfies the constraint analytically. In practice, the constraint is satisfied to the precision of the automatic differentiation and floating-point arithmetic, which is typically near machine epsilon; far better than the $\mathcal{O}(h^p)$ violations common in grid-based methods. Boundary layers and current sheets, which in traditional methods demand adaptive mesh refinement, are handled implicitly by the network's capacity to represent sharp gradients through its nonlinear activation functions. Furthermore, PINNs naturally accommodate inverse problems: unknown physical parameters such as the resistivity $\eta$ or viscosity $\mu$ can be included as trainable variables and inferred simultaneously with the solution from partial observations~\citep{RAISSI2019686}. Karniadakis and collaborators subsequently articulated a broader vision for \emph{physics-informed machine learning} in an influential review~\citep{Karniadakis2021}, positioning PINNs within a larger framework that encompasses Bayesian inference, operator learning, and the systematic integration of physical constraints into data-driven models. This review helped establish a common vocabulary and set of benchmarks for the field, and it explicitly identified MHD and plasma physics as high-value application domains. Despite these strengths, PINNs exhibit well-documented limitations that become acute for MHD problems. First, each trained PINN solves a \emph{single} PDE instance: changing the initial conditions, boundary geometry, or physical parameters requires retraining from scratch. For applications demanding many evaluations, e.g., parametric studies, uncertainty quantification, real-time control, this per-instance cost is prohibitive. Second, the multi-term loss function~\eqref{eq:pinn_loss} creates a challenging multi-objective optimization landscape. Wang, Teng, and Perdikaris~\citep{Wang2021} demonstrated that gradient pathologies arise when the PDE residual loss and the data/boundary losses have disparate magnitudes or convergence rates, leading to slow training or convergence to poor local minima. For MHD, where the governing equations involve multiple coupled fields ($\rho$, $\mathbf{v}$, $\mathbf{B}$, $p$) with different characteristic scales, these balancing difficulties are particularly severe. Third, PINNs struggle with high-frequency and multiscale solutions~\citep{Wang2021}, precisely the regime relevant to MHD turbulence and high-Lundquist-number dynamics. These limitations motivated a conceptual shift: rather than training networks to approximate individual solutions, one should train them to approximate the \emph{operator} that maps inputs (initial conditions, forcing, parameters) to solutions. 

\section{The Broader Landscape of AI Methods for PDEs and MHD} 

\noindent While PINNs established the foundational paradigm of embedding physical laws into neural network training, the subsequent development of AI methods for PDE solving has produced a diverse ecosystem of architectures. Each carries distinct strengths and limitations relevant to the MHD equations. Before focusing on the Fourier neural operator family that forms the backbone of the methods detailed in \S\ref{sec:neural_op}--\S\ref{sec:diff_model}, it is important to map this broader landscape and identify the opportunities each approach offers for magnetised fluid dynamics. 

\noindent \textbf{Operator learning beyond Fourier neural operators.} The DeepONet architecture of Lu, Jin, and Karniadakis~\cite{lu2021deepONet} provides an alternative operator-learning framework grounded in the universal approximation theorem for nonlinear continuous operators established by Chen and Chen~\cite{TiaChe_1995}. DeepONet decomposes the learned operator into a branch network, which encodes the input function (e.g., initial conditions or forcing), and a trunk network, which encodes the output coordinates (spatial and temporal locations). This factored structure permits evaluation at arbitrary output points without retraining, a property shared with the FNO's discretisation invariance but achieved through a fundamentally different mechanism. Lu et al.\ demonstrated DeepONet on a range of PDE systems including advection--diffusion, reaction--diffusion, and the Euler equations of gas dynamics~\cite{lu2021deepONet}. For MHD, DeepONet's flexibility in handling non-uniform evaluation points makes it a natural candidate for fusion applications involving complex tokamak geometries where uniform Fourier grids are impractical. Physics-informed extensions of DeepONet, in which the PDE residual is incorporated into the training loss analogously to PINO, have been developed by Wang, Wang, and Perdikaris~\cite{Sifan2021}, demonstrating improved generalisation and reduced data requirements on multi-physics problems.

A systematic comparison of FNO and DeepONet architectures on canonical benchmarks was presented by Lu et al.~\cite{LuXu_2022}, revealing that FNOs tend to excel on problems with periodic boundary conditions and smooth solutions, while DeepONet offers advantages on irregular domains and problems with localised features, precisely the regime relevant to current-sheet formation in MHD. 

\noindent \textbf{Graph neural networks for unstructured meshes.} Many production MHD codes, including JOREK~\cite{Hoelzl2021}, M3D-C$^1$~\cite{Krebs2017}, and finite-element formulations of GRMHD, operate on unstructured or adaptive meshes that are poorly suited to the uniform-grid assumption underlying Fourier-based architectures. Graph neural networks (GNNs), which define message-passing operations on arbitrary graph topologies, offer a natural alternative. Pfaff et al.\ introduced MeshGraphNets~\cite{pfafflearning}, demonstrating that GNNs can learn to simulate complex fluid dynamics, including cloth, incompressible flow around obstacles, and compressible flow, on triangular meshes with accuracy comparable to finite-element solvers at substantially reduced computational cost. The key architectural idea is to represent the simulation mesh as a graph, with nodes at mesh vertices and edges connecting neighbouring vertices, and to learn update rules that respect the local connectivity structure. 

For MHD, GNN architectures could be particularly valuable for problems involving adaptive mesh refinement around current sheets and reconnection sites, where the mesh topology changes dynamically during the simulation. Lino et al.\ extended graph-based approaches to multi-scale fluid simulations~\cite{LinoMa_2022}, demonstrating that hierarchical graph structures can capture cross-scale interactions analogous to the multi-grid methods used in traditional MHD solvers. The principal limitation of GNNs for turbulence is their local message-passing structure: capturing long-range correlations, such as those mediated by Alfv\'en waves propagating along the guide field, requires either very deep networks or explicit long-range edges, both of which increase computational cost.

\noindent \textbf{Transformer architectures for long-range interactions.} The attention mechanism central to transformer architectures provides a natural framework for capturing long-range spatial correlations without the depth limitations of local message-passing. In MHD, Alfv\'en waves mediate non-local coupling along magnetic field lines over distances comparable to the system size, making long-range attention a physically motivated architectural choice. Li et al.\ developed the Factorised Transformer (FactFormer)~\cite{li2023scalable}, which decomposes the full attention matrix into axial components to reduce the quadratic cost of standard attention from $\mathcal{O}(N^2)$ to $\mathcal{O}(N^{4/3})$ for three-dimensional problems, enabling application to high-resolution PDE solving. Hao et al.\ introduced the General Neural Operator Transformer (GNOT)~\cite{hao2023gnot}, which combines cross-attention between input functions and query coordinates with self-attention among query points, achieving competitive accuracy on multi-physics problems including Navier--Stokes and convection--diffusion systems. For 3D MHD turbulence, where the anisotropic cascade ($k_{\parallel} \sim k_{\perp}^{2/3}$) creates elongated structures along the guide field, transformer architectures with anisotropic attention patterns, attending preferentially along field-aligned directions, represent a promising but as yet unexplored direction. 

\noindent \textbf{Structure-preserving and equivariant networks.} A fundamental concern in applying neural networks to MHD is the preservation of physical invariants: the solenoidal constraint $\nabla \cdot \mathbf{B} = 0$, energy conservation in the ideal limit, magnetic helicity conservation, and the symmetries of the governing equations under rotations and translations. Greydanus, Dzamba, and Sprague introduced Hamiltonian Neural Networks (HNNs)~\cite{greydanus2019hamiltonian}, which parameterise the Hamiltonian of a system and derive the dynamics via Hamilton's equations, thereby exactly conserving the learned energy. Extensions to dissipative systems and to field theories have been developed by several groups. For MHD, the ideal invariants (total energy, cross-helicity, and magnetic helicity) play a central role in constraining the cascade dynamics, and architectures that conserve these quantities by construction could dramatically improve long-time prediction accuracy. 

Brandstetter et al.\ introduced Clifford Neural Layers~\cite{brandstetterclifford} that encode geometric algebra operations, including the distinction between divergence-free and curl-free vector fields, directly into the network architecture. Such layers could enforce $\nabla \cdot \mathbf{B} = 0$ as a hard architectural constraint rather than a soft penalty in the loss function, analogous to the constrained-transport approach in grid-based codes. The development of neural network layers that implement a discrete exterior calculus, preserving the de~Rham complex structure that underlies the mathematical consistency of Maxwell's equations, remains an open and high-impact research direction for AI-driven MHD.

\noindent \textbf{Data-driven turbulence closures.} Large-eddy simulation (LES) of MHD turbulence requires sub-grid-scale (SGS) models that represent the effect of unresolved scales on the resolved flow. Traditional SGS models, such as Smagorinsky-type eddy viscosity and dynamic models, are known to perform poorly for MHD, where the magnetic field introduces additional SGS stresses (the Maxwell SGS stress tensor) and the cascade dynamics differ qualitatively from hydrodynamic turbulence. Machine learning offers a systematic alternative: neural networks trained on filtered DNS data can learn SGS closures that capture the correct energy transfer between resolved and unresolved scales. Duraisamy, Iaccarino, and Xiao provided an influential framework for data-driven turbulence modelling encompassing both Reynolds-averaged and LES closures~\cite{dura_2019}. Beck, Flad, and Munz demonstrated that convolutional neural networks can learn SGS models for compressible turbulence that outperform traditional closures in both \emph{a priori} and \emph{a posteriori} tests~\cite{BECK2019108910}. Extending these approaches to MHD, where the SGS model must simultaneously close the momentum and induction equations while preserving the solenoidal constraint, is an active and promising direction. Grete et al.~\cite{Grete_2015} analysed the structure of the SGS stress tensors in MHD turbulence, providing the physical foundation on which data-driven closures can be built. The combination of learned SGS models with the operator-learning frameworks described in \S\ref{sec:diff_model} could dramatically reduce the resolution requirements of production MHD simulations while preserving the correct cascade dynamics. 

\noindent \textbf{Latent-space and reduced-order methods.} An alternative to learning full-resolution solution operators is to first compress the high-dimensional MHD state into a low-dimensional latent representation using an autoencoder, and then learn the dynamics in latent space using neural ODEs or recurrent architectures. This approach exploits the empirical observation that turbulent flows, despite their high-dimensional state space, often evolve on low-dimensional attractors. Geneva and Zabaras demonstrated latent-space modelling for turbulent flows using transformers operating in a learned latent space~\cite{geneva2022transformers}, showing that such models can capture long-time statistics with orders-of-magnitude fewer degrees of freedom than the full simulation. For MHD, where the state vector includes both velocity and magnetic fields with coupled dynamics, the design of latent representations that preserve the cross-field correlations (encoded in the Els\"asser variables $z^{\pm} = v \pm b$) and the solenoidal constraint is a non-trivial challenge that has not yet been systematically addressed. 

These diverse approaches are not mutually exclusive; indeed, the most promising path forward likely involves their hybridisation. The DINOs framework reviewed in \S\ref{sec:diff_model}, which combines a physics-informed neural operator with a score-based diffusion model, exemplifies this hybrid philosophy. Future architectures may combine graph-based spatial representations with Fourier-based spectral processing, transformer attention for long-range Alfv\'enic correlations, and equivariant layers for constraint preservation, all within a single end-to-end trainable framework. The challenge is to assemble these components in a way that respects the specific mathematical structure of the MHD equations while remaining computationally tractable at the scales demanded by astrophysical and fusion applications.

\section{Neural Operators for MHD} 
\label{sec:neural_op}

The operator learning paradigm was crystallized by the FNO of Li et al.~\citep{Li2020}, developed in the group of Anima Anandkumar at Caltech. The FNO learns a mapping $\mathcal{G}_\theta: \mathcal{A} \to \mathcal{U}$ between function spaces, for instance, from the space of initial conditions $\mathcal{A}$ to the space of solutions $\mathcal{U}$ at a later time. The architecture parameterizes this mapping through a sequence of \emph{Fourier layers}, each of which applies a linear transformation in the frequency domain followed by a pointwise nonlinearity:

\begin{equation}
v^{(\ell+1)}(\mathbf{x})
=
\sigma\!\left(
W^{(\ell)}\,v^{(\ell)}(\mathbf{x})
+
\mathcal{F}^{-1}\!\left[
R^{(\ell)} \cdot \mathcal{F}\!\left[v^{(\ell)}\right]
\right]\!(\mathbf{x})
\right),
\label{eq:fno_layer}
\end{equation}

where $\ell$ indexes the Fourier layer, $\mathcal{F}$ denotes the discrete Fourier transform, $R^{(\ell)} \in \mathbb{C}^{d_v \times d_v \times k_{\max}}$ is a learnable complex-valued weight tensor acting on the retained Fourier modes $|k| \leq k_{\max}$, $W^{(\ell)} \in \mathbb{R}^{d_v \times d_v}$ is a pointwise linear map, and $\sigma$ is a pointwise nonlinear activation function.

A crucial property of this architecture is \emph{discretization invariance}: once trained, the operator can be evaluated on grids of different resolution, since the underlying map is defined between continuous function spaces rather than between fixed-dimensional vectors. The restriction to a finite number of low-frequency Fourier modes, however, introduces an intrinsic spectral bias, a design choice that proves consequential for turbulent flows, as we shall see. The FNO demonstrated order-of-magnitude speedups over traditional solvers on benchmark problems including the Navier--Stokes equations in the turbulent regime~\citep{Li2020}. However, as a purely data-driven method, the FNO requires large volumes of high-fidelity training data and offers no guarantee that its predictions satisfy the governing equations. Extrapolation beyond the training distribution, e.g., to higher Reynolds numbers, different forcing configurations, or longer time horizons, can produce physically inconsistent results. To address this, Li et al.\ introduced the \emph{physics-informed neural operator} (PINO)~\citep{Li2024PINO}, which augments the data-driven training of the FNO with a PDE residual loss analogous to that used in PINNs: \begin{equation} \mathcal{L}_{\text{PINO}} = \lambda_{\text{data}}\,\mathcal{L}_{\text{data}} + \lambda_{\text{IC}}\,\mathcal{L}_{\text{IC}} + \lambda_{\text{PDE}}\,\mathcal{L}_{\text{PDE}}. \label{eq:pino_loss} \end{equation} 

Here $\mathcal{L}_{\text{data}}$ measures the discrepancy between the operator's output and reference solutions from numerical simulations, $\mathcal{L}_{\text{IC}}$ enforces agreement with initial conditions, and $\mathcal{L}_{\text{PDE}}$ penalizes the PDE residual of the operator's output, computed via automatic differentiation through the neural operator.

The systematic application of PINOs to the MHD equations was pioneered by Rosofsky and Huerta~\citep{Rosofsky2023PINO}. This work confronted the specific challenges that distinguish MHD from the scalar and incompressible-flow problems on which neural operators had previously been validated, and it established both the promise and the limits of deterministic operator surrogates for magnetized fluid dynamics. Rosofsky and Huerta considered two-dimensional, incompressible, resistive MHD on a doubly periodic domain, governed by coupled evolution equations for the vorticity $\omega$ and the magnetic vector potential $A$ (from which the magnetic field is derived as $\mathbf{B} = \nabla\times A\,\hat{\mathbf{z}}$, guaranteeing $\nabla\cdot\mathbf{B}=0$ to machine precision). Even in this simplified two-dimensional setting, the system retains the essential physics of MHD turbulence: nonlinear coupling between velocity and magnetic fields through the Lorentz force, dual cascades of kinetic and magnetic energy, current-sheet formation, and magnetic reconnection events. The complexity substantially exceeds that of the canonical benchmarks, namely, Burgers' equation, Darcy flow, Kolmogorov flow, on which neural operators had been tested. Their approach employed a tensor-factorized FNO architecture trained with the composite physics-informed loss, see Equation~\eqref{eq:pino_loss}. The PDE residual term incorporates the full system of MHD equations. Training data were generated using the spectral solver \textsc{Dedalus}~\citep{Burns2020}, with initial conditions drawn from Gaussian random fields to ensure diversity in the training ensemble. The trained PINO achieved high accuracy for flows in the laminar and mildly turbulent regimes, up to $Re \approx 250$, faithfully reproducing the spatio-temporal evolution of both velocity and magnetic fields across grid resolutions up to $256^2$. However, the work also revealed a critical limitation. As the Reynolds number exceeded approximately 500, the PINO's accuracy degraded markedly, with increasing mean-squared errors concentrated in the small-scale magnetic field structures~\citep{Rosofsky2023PINO}. This degradation is a direct consequence of the FNO's spectral truncation: the Fourier layers retain only a fixed number of low-frequency modes, and any energy residing at higher wavenumbers is discarded. In laminar flows, where the energy spectrum decays rapidly, this truncation is benign. In turbulent flows, where energy cascades populate a broad range of wavenumbers, the truncation amounts to an uncontrolled low-pass filter that systematically suppresses the fine-scale structures that carry an increasing fraction of the system's energy. The magnetic field, whose spectrum typically extends to higher wavenumbers than the velocity field due to the formation of thin current sheets, is particularly vulnerable to this spectral bias. In complementary work, Rosofsky and Huerta~\citep{Rosofsky2023MLST} investigated the broader applicability of physics-informed neural operators to systems of conservation laws, examining how the interplay between data-driven and physics-informed training affects accuracy, stability, and generalization. This study provided practical guidance on loss function design, training strategies, and architecture choices for complex PDE systems, contributing to the methodological foundations on which subsequent MHD applications have been built. The results of Rosofsky and Huerta thus established a clear picture: PINOs are powerful surrogates for MHD in regimes where the solution is spectrally compact, but they are fundamentally limited by their deterministic, spectrally truncated architecture when confronted with the broadband energy spectra of developed turbulence. Overcoming this limitation required a qualitatively different approach, one capable of generating the high-frequency content that the neural operator, by construction, cannot represent. 

\section{Diffusion-Integrated Neural Operators for Turbulent MHD} 
\label{sec:diff_model}

\begin{figure}[!htbp]
    \centering
    \includegraphics[width=\textwidth]{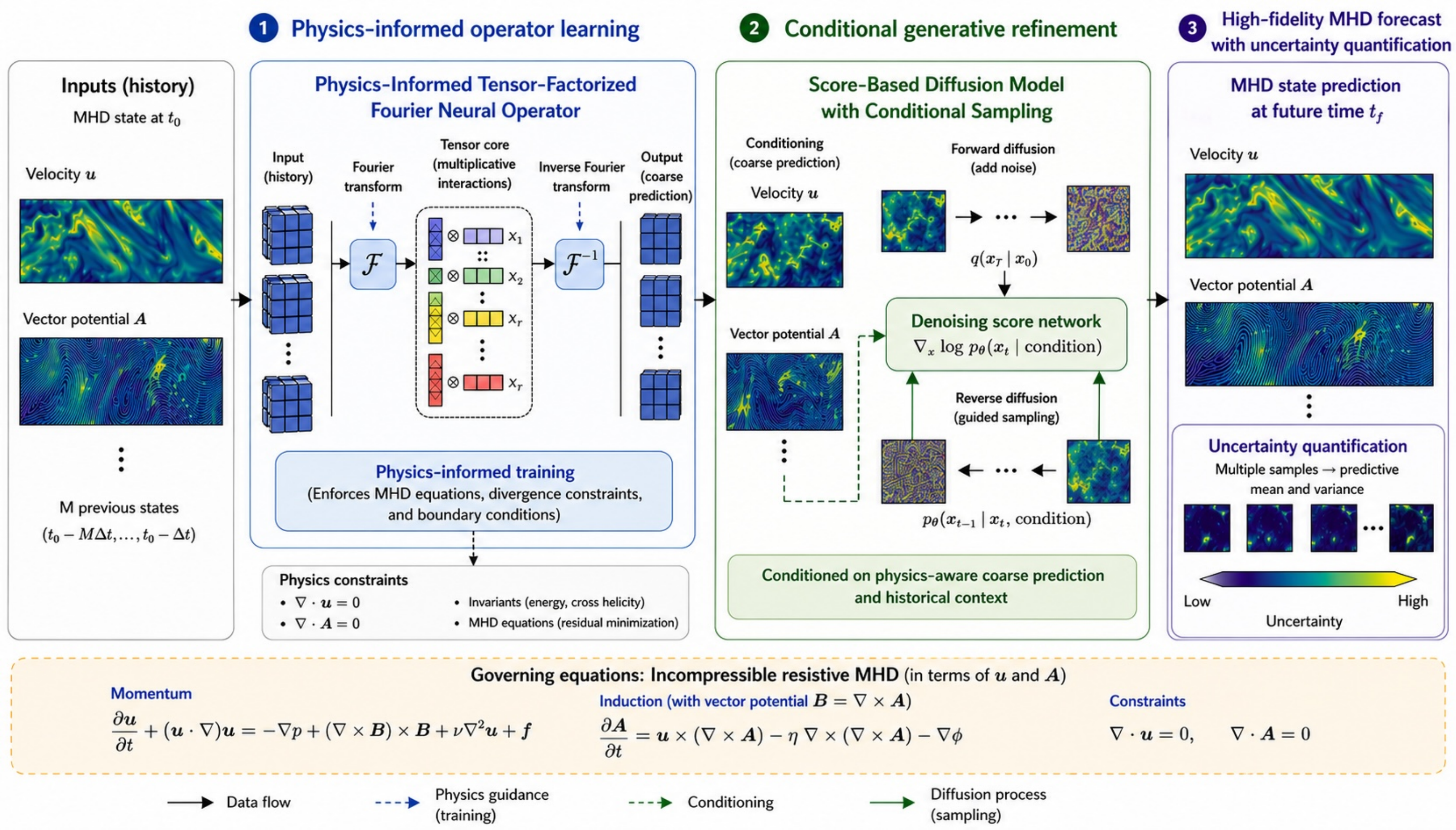}
    \caption{Schematic of the DINOs (Diffusion-Integrated Neural Operators) framework~\cite{Kacmaz2025}. \textbf{Stage~1:} A physics-informed tensor-factorized Fourier neural operator (PINO) predicts the large-scale evolution of the velocity field $\mathbf{u}$ and magnetic vector potential $A$, capturing modes up to a cutoff wavenumber $k_{\max}$. \textbf{Stage~2:} A conditional score-based diffusion model, built on a UNet backbone, acts as a stochastic corrector that restores the unresolved high-wavenumber content ($k > k_{\max}$). The hybrid architecture reduces the relative $L^2$ error from $25.5\%$ (PINO only) to $10.3\%$ at $\mathrm{Re}=1000$ and from $39.1\%$ to $20.5\%$ at $\mathrm{Re}=10{,}000$~\cite{Kacmaz2025}. The relative $L^{2}$ error is measured against ground-truth solutions computed with the pseudospectral solver \textsc{Dedalus}~\cite{Burns2020}.}
    \label{fig:dinos}
\end{figure}

The resolution of this spectral bottleneck was achieved by 
Kacmaz, Huerta, and Haas~\citep{Kacmaz2025}, who introduced a hybrid framework, termed \emph{diffusion-integrated neural operators} (DINOs), that combines a physics-informed neural operator with a conditional score-based diffusion model (see Figure~\ref{fig:dinos}).

This work, published in \emph{Machine Learning: Science and Technology} in 2025, represents the first application of a hybrid operator-diffusion architecture to MHD and achieves state-of-the-art accuracy in turbulent regimes previously inaccessible to deterministic surrogates. The intellectual motivation for the hybrid approach rests on a precise diagnosis of complementary failure modes. As established by Rosofsky and Huerta, the PINO excels at capturing the large-scale, low-frequency dynamics of MHD flows but fails to represent high-wavenumber content due to its spectral truncation. 

Conversely, score-based diffusion models~\citep{Ho2020,Song2021}, i.e., generative models that learn to reverse a gradual noising process by estimating the score function $\nabla_{\mathbf{x}}\log p_t(\mathbf{x})$ of the perturbed data distribution, are powerful generators of fine-scale structure but lack the inductive bias to produce temporally coherent, physically consistent sequences when applied monolithically to PDE dynamics. As noted by Guo et al.~\citep{Guo2025}, standard diffusion frameworks treat prediction as a conditional generation task without explicitly modeling temporal evolution between adjacent states, limiting their ability to generate long, coherent sequences. Furthermore, tasking a single diffusion model with learning the full joint distribution of velocity and magnetic fields risks learning the marginal distributions of each field independently, failing to capture the cross-field correlations that govern MHD evolution. The DINOs framework resolves these difficulties through a principled two-stage architecture inspired by the hybrid operator-diffusion methodology of Oommen et al.~\citep{Oommen2025}. In the first stage, a PINO, built on a tensor-factorized FNO backbone with 8 Fourier layers, 32 latent channels, and canonical polyadic tensor decomposition of the spectral weights~\citep{Kossaifi2019,Kolda2009}, is trained with the composite physics-informed loss incorporating the full 2D incompressible resistive MHD equations: 

\begin{align} 
\rho \left( \frac{\partial \mathbf{v}}{\partial t} + (\mathbf{v}\cdot\nabla)\mathbf{v}\right) &= -\nabla P + \frac{1}{4\pi}(\nabla\times\mathbf{B})\times\mathbf{B} + \nu\nabla^2\mathbf{v}, \label{eq:mhd_momentum}\\ \frac{\partial A}{\partial t} + (\mathbf{u}\cdot\nabla)A &= \eta\nabla^2 A, \label{eq:mhd_induction} \end{align}

subject to $\nabla\cdot\mathbf{v}=0$ and $\nabla\cdot\mathbf{B}=0$, the latter enforced exactly by evolving the magnetic vector potential $A$ rather than $\mathbf{B}$ directly. This PINO produces a smooth, spectrally limited approximation of the true solution, accurate in the low-wavenumber content but missing the high-frequency turbulent structures. In the second stage, a conditional score-based diffusion model acts as a stochastic corrector. Built on a UNet backbone~\citep{Ronneberger2015} with a base dimension of 128 and six down/up-sampling stages, the diffusion model is conditioned on the PINO output by concatenating it with the noisy state along the channel dimension. The model is trained via denoising score matching~\citep{Vincent2011} to learn the conditional score $\nabla_{\mathbf{x}}\log p_t(\mathbf{x}|\mathbf{y})$, where $\mathbf{y}$ is the PINO prediction. During inference, the reverse-time SDE, integrated from the noise level $t=T$ back to $t=0$, is solved 

\begin{equation} d\mathbf{x} = \left[\mathbf{f}(\mathbf{x},t) - g(t)^2\,\nabla_{\mathbf{x}}\log p_t(\mathbf{x}|\mathbf{y})\right]dt + g(t)\,d\bar{\mathbf{w}} \label{eq:reverse_sde} 
\end{equation} 

where time runs backward from $T$ (pure noise) to $0$ (clean sample). The drift coefficient
\begin{equation}
\mathbf{f}(\mathbf{x},t) = -\frac{1}{2}\beta(t)\,\mathbf{x}\,,
\end{equation}

and diffusion coefficient

\begin{equation}
g(t) = \sqrt{\beta(t)}\,,
\end{equation}

are inherited from the forward variance-preserving SDE

\begin{equation}
d\mathbf{x} = -\frac{1}{2}\beta(t)\,\mathbf{x}\,dt
+ \sqrt{\beta(t)}\,d\mathbf{w}\,,
\label{eq:forward_sde}
\end{equation}

which gradually corrupts data into Gaussian noise over the interval

\begin{equation}
t \in [0,T]\,.
\end{equation}

The conditional score

\begin{equation}
\nabla_{\mathbf{x}} \log p_{t}(\mathbf{x} \mid \mathbf{y})
\end{equation}

is approximated by the trained UNet

\begin{equation}
\mathbf{s}_{\theta}(\mathbf{x},t,\mathbf{y})\,,
\end{equation}

and $\bar{\mathbf{w}}$ denotes a standard Wiener process running backward in time. This division of labor is not merely a convenience but a necessity: as analyzed by Oommen et al.~\citep{Oommen2025}, the score function's gradient is inherently weaker at low wavenumbers, making the diffusion model more effective at refining high-frequency components than at generating accurate large-scale structure from scratch. Kacmaz, Huerta, and Haas trained and evaluated their framework on a comprehensive dataset of high-fidelity MHD simulations generated with the spectral solver \textsc{Dedalus}~\citep{Burns2020}, spanning Reynolds numbers $Re \in \{100, 250, 500, 750, 1000, 3000, 10\,000\}$ with 1000 simulations per Reynolds number. The training was performed at leadership-scale computing facilities: the PINO was trained on AMD MI250x GPUs on the Frontier supercomputer at the Oak Ridge Leadership Computing Facility, while the diffusion model was trained on NVIDIA H100 GPUs on the NCSA DeltaAI cluster. The results demonstrate a consistent and substantial improvement over the PINO-only baseline across all turbulent regimes. At $Re = 1000$, the time-averaged relative $L^2$ error drops from 25.5\% (PINO only) to 10.3\% (DINOs), a nearly threefold reduction. At $Re = 3000$, the error decreases from 32.7\% to 15.9\%. Even at the most extreme case tested, $Re = 10\,000$, the framework halves the error from 39.1\% to 20.5\%, and most importantly, produces physically plausible flow fields where the PINO alone generates unphysically smooth states that bear little resemblance to the true turbulent dynamics. Beyond pointwise error metrics, the framework's fidelity is most convincingly demonstrated through spectral analysis. At $Re = 1000$ and $Re = 3000$, the DINOs framework accurately reconstructs the kinetic and magnetic energy spectra across the full wavenumber range, correcting the severe spectral deficit exhibited by the PINO-only model at high wavenumbers. The magnetic field spectrum, which is the first to degrade under spectral truncation due to the formation of thin current sheets, is restored with particular effectiveness. At $Re = 10\,000$, the DINOs framework remains the first surrogate model capable of recovering the high-wavenumber evolution of the magnetic field, preserving large-scale morphology and enabling statistically meaningful predictions even where point-wise accuracy necessarily decreases. The framework also exhibits promising generalization properties. Cross-regime experiments showed that while models trained at $Re = 1000$ cannot generalize to the qualitatively different laminar regime at $Re = 100$ (producing errors exceeding 85\%), they transfer effectively to nearby turbulent regimes: applied to $Re = 900$ test data, the diffusion model successfully corrected the parametric mismatch in the PINO prediction, reducing the error from approximately 40\% to 13\%. This suggests that the diffusion component possesses a robust corrective capability that could be exploited through fine-tuning strategies, reducing the cost of developing surrogate models for new parameter regimes. 

\section{Conclusions} 
\label{sec:end}

The progression reviewed here, from PINNs to PINOs to diffusion-integrated neural operators, reflects a deepening engagement between the AI and MHD communities, with each methodological advance driven by specific physical and mathematical demands of the MHD equations. The PINN framework of Karniadakis and collaborators established that neural networks could be trained to satisfy PDEs directly, but was limited to single instances. The PINO framework, building on the FNO architecture of Anandkumar and collaborators, extended this to operator learning across families of problems, and Rosofsky and Huerta demonstrated both its power and its spectral limitations for MHD. The DINOs framework of Kacmaz, Huerta, and Haas resolved the spectral bottleneck by combining the operator's physically constrained low-frequency predictions with a diffusion model's capacity to generate high-frequency turbulent structure, achieving state-of-the-art results across a broad range of Reynolds numbers. Several open challenges will shape the next phase of development. The DINOs framework has been demonstrated for two-dimensional, incompressible MHD at unit magnetic Prandtl number $\mathrm{Pr}_m = \nu/\eta = 1$. Extension to three dimensions, where anisotropy, Alfv\'{e}nic dynamics, and qualitatively different cascade phenomenology emerge, and to $\mathrm{Pr}_m \neq 1$ regimes relevant to both astrophysical ($\mathrm{Pr}_m \ll 1$ in stellar interiors) and laboratory ($\mathrm{Pr}_m \gg 1$ in some fusion plasmas) settings remains an active direction of research~\cite{Kacmaz2025}.

Long-time stability of autoregressive predictions, generalization across qualitatively different physical regimes, and the treatment of compressible MHD with shocks and discontinuities are further challenges that will require both architectural innovations and new training strategies. Parallel developments reinforce the trajectory described here. Deep reinforcement learning has been used for real-time magnetic confinement control in tokamaks~\cite{Degrave2022}. Operator learning frameworks such as DeepONet~\cite{lu2021deepONet} offer alternative architectures with complementary strengths. The hybrid operator-diffusion methodology introduced by Oommen et al.~\cite{Oommen2025} for fluid dynamics has now been shown by Kacmaz, Huerta, and Haas to extend to the richer physics of MHD, suggesting broad applicability across multi-scale problems in computational physics. Looking ahead, the convergence of physics-constrained AI with exascale and post-exascale computing charts a clear trajectory. The integration of learned sub-grid-scale closures within production MHD solvers offers a near-term path to reduce grid requirements by orders of magnitude while preserving cascade dynamics~\cite{dura_2019,BECK2019108910}.

Structure-preserving architectures~\cite{greydanus2019hamiltonian,brandstetterclifford} that enforce the solenoidal constraint $\nabla \cdot \mathbf{B} = 0$ and conserve ideal invariants by construction, rather than by penalization, provide a principled route toward long-time stability in autoregressive rollouts. Graph neural networks~\cite{pfafflearning} and transformer-based operator learners~\cite{li2023scalable,hao2023gnot} supply complementary architectures suited to the unstructured meshes and long-range Alfv\'{e}nic correlations that characterize three-dimensional MHD in complex geometries. Hybrid workflows, in which neural operators accelerate the most expensive solver components while the host code retains control of conservation laws and shock capturing, could extend direct numerical simulation to Lundquist numbers $S \sim 10^{8}$--$10^{9}$, narrowing the gap to the $S \sim 10^{10}$ regime relevant to solar coronal dynamics and ITER-scale fusion plasmas.

On a longer horizon, fault-tolerant quantum algorithms for sparse linear systems~\cite{Harrow2009}, together with near-term variational quantum linear solvers~\cite{Bravo2023}, may eventually provide additional leverage for the implicit solves central to resistive MHD, although practical quantum advantage will require substantial advances in error correction and qubit connectivity beyond current hardware capabilities. The work reviewed here has established that AI methods can engage meaningfully with the full complexity of MHD turbulence, not merely reproducing smooth, large-scale features, but recovering the broadband spectral content, intermittent structures, and non-Gaussian statistics that define the turbulent state. The challenge ahead is to extend these methods to the three-dimensional, multi-physics, multi-scale regimes that characterize the most demanding applications in astrophysics, fusion science, and space weather prediction.

\begin{acknowledgement}
E.H. acknowledges insightful conversations with Semih Kacmaz, Shawn Rosofsky and Roland Haas. He also acknowledges support from NSF grants OAC-2514142 and OAC-2209892. This work was partially supported by the U.S. Department of Energy under contract No. DE-AC02-06CH11357, including funding from the Office of Advanced Scientific Computing Research's Diaspora project and the Laboratory Directed Research and Development program.  
\end{acknowledgement}

\bibliographystyle{plain}
\bibliography{references}

\end{document}